\documentclass[12pt,showpacs,preprintnumbers,superscriptaddress,amsmath,amssymb,nofootinbib]{revtex4}
\usepackage{graphicx}
\usepackage{dcolumn}
\usepackage{bm}
\usepackage{amssymb}
\usepackage{amsmath}
\usepackage{epsfig}    
\usepackage{color}
\usepackage{hhline}

\def\be{\begin{equation}}
\def\ee{\end{equation}}
\newcommand{\bea}{\begin{eqnarray}}
\newcommand{\eea}{\end{eqnarray}}
\newcommand{\nn}{\nonumber}

\numberwithin{equation}{section}

\begin{document}

\title{3.55 keV X-ray Line Interpretation in Radiative Neutrino Model}
\preprint{KIAS-P14036}
\preprint{SU-HET-08-2014}

\author{Hiroyuki Ishida}
\email{ishida@riko.shimane-u.ac.jp}
\affiliation{Graduate School of Science and Engineering, Shimane University, Matsue, Shimane 690-8504, Japan}
\author{Hiroshi Okada}
\email{hokada@kias.re.kr}
\affiliation{School of Physics, KIAS, Seoul 130-722, Korea}

\begin{abstract}
We discuss the 3.55 keV X-ray line anomaly reported by XMN-Newton X-ray observatory
using data of various galaxy clusters and Andromeda galaxy in a radiative neutrino model,
in which the mixing between the active neutrino and the dark matter is generated at two-loop level after the spontaneous breaking of $Z_2$ symmetry.
It might provide us a natural explanation of its tiny mixing ${\cal O}(10^{-10})$, which is observed by their experiments.
Such an Abelian discrete symmetry plays a crucial role in differentiating the TeV scale Majorana field from our dark matter, whose mass is expect to be around 7.1 keV. 

\end{abstract}
\maketitle
\newpage

\section{Introduction}
The standard model (SM) has been finalized by the discovery of the Higgs boson.
On the other hand, 
so many unsolved problems in the particle physics and cosmology still remain.
The origin of the dark matter (DM) is 
one of the biggest issues for such mysteries.
Its properties are well known by previous experiments that tell us 
 an electrically neutral particle, a warm or cold one, and longer lifetime than the universe,
if DM is constituted by an elementary particle.
It should be noted that 
the last one does not mean that DM must be a completely stable particle.
One of the plausible possibilities for such an unstable DM is a gauge singlet fermion.
If the magnitude of its interaction is tiny, 
the lifetime can be longer than the age of the universe.

Recently, two different groups~\cite{Bulbul:2014sua,Boyarsky:2014jta} report 
a signal of an anomalous X-ray line which has an energy of about $3.55~{\rm keV}$.
A lot of literature has recently arisen around the subject after this announcement~\cite{Ishida:2014dlp, Finkbeiner:2014sja, 
Higaki:2014zua, Jaeckel:2014qea,Lee:2014xua, Kong:2014gea,
 Frandsen:2014lfa, Baek:2014qwa, Cline:2014eaa, Modak:2014vva,
 Babu:2014pxa, Queiroz:2014yna, Demidov:2014hka, Ko:2014xda,
 Allahverdi:2014dqa,  Kolda:2014ppa, Cicoli:2014bfa, Dudas:2014ixa, Choi:2014tva, Okada:2014zea, Chen:2014vna, Conlon:2014xsa, Robinson:2014bma, Liew:2014gia, Chakraborty:2014tma, Tsuyuki:2014aia}.
This signal may come from a radiative decay of the gauge singlet fermion 
and constrains the mass of the singlet fermion as about $7.1~{\rm keV}$ 
and the mixing with the ordinary neutrinos as ${\cal O}(10^{-10})$.
To simply realize these small mass and mixing angle, 
the parameters of the Majorana mass and tree level coupling 
should be artificially made tiny.
However, another solution can be considered 
when the scalar sector is extended by some fields with additional symmetries that forbid tree level contributions.
Such symmetries often assure the stability of DM. Hence the
observed neutrino masses can be generated through a radiative correction 
with $\mathcal{O}(1)$ couplings 
as well as the mixing angle is successfully suppressed.

This paper is organized as follows.
In Sec.~II, we show our model building including neutrino mass.
In Sec.~III, we discuss the DM nature. We conclude in Sec.~VI.

\section{The Model}
In this section, we devote the detail of our model.
First, we show the charge assignments for each field 
and introduce the properties of them.
Second, we analyze the influence of the additional scalars to 
the so-called Peskin-Takeuchi parameters.
Finally, we discuss the neutrino mass generation 
by one of the radiative seesaw models, Ma-model~\cite{Ma:2006km, Hehn:2012kz}.

\subsection{Model setup}

\begin{table}[thbp]
\centering {\fontsize{10}{12}
\begin{tabular}{|c||c|c|c|c|}
\hline Particle & $L_L$ & $ e_{R} $ & $N_R$   & $X_R$ 
  \\\hhline{|=#=|=|=|=|$}
$(SU(2)_L,U(1)_Y)$ & $(\bm{2},-1/2)$ & $(\bm{1},-1)$ & $(\bm{1},0)$  & $(\bm{1},0)$
\\\hline
$Z_2$ & $-$ & $-$ &  $+$ & $-$   \\\hline
$Z'_2$ & $+$ & $+$ &  $+$ & $-$   \\\hline
\end{tabular}%
} \caption{The particle contents for fermions, where $L_L$ and $e_R$ have three generations, on the other hand, $N_R$ and $X_R$ are introduced as one family. $X_R$ is our DM to explain the X-ray line signal.} 
\label{tab:1}
\end{table}

\begin{table}[thbp]
\centering {\fontsize{10}{12}
\begin{tabular}{|c||c|c|c|c|}
\hline Particle   & $\Phi$   & $\eta_i\ (i=1\,,2)$   & $\chi^+$   & $\chi^0$ 
  \\\hhline{|=#=|=|=|=|}
$(SU(2)_L,U(1)_Y)$ & $(\bm{2},1/2)$  & $(\bm{2},1/2)$  & $(\bm{1},1)$  & $(\bm{1},0)$ \\\hline
$ Z_2$  & $+$ & $-$ & $+$  & $+$    \\\hline
$ Z'_2$  & $+$ & $+$ & $-$  & $-$   \\\hline
\end{tabular}%
} \caption{The particle contents and the charges for scalars. }
\label{tab:2}
\end{table}

We discuss the one-loop induced radiative neutrino model 
containing the DM candidate with the mass at keV range. 
The particle contents are shown in Tab.~\ref{tab:1} and Tab.~\ref{tab:2}. 
We add two $SU(2)_L$ singlet Majorana fermions $N_R$ and $X_R$, 
where $X_R$ is assumed to be the DM candidate.
For new bosons, we introduce two $SU(2)_L$ doublet scalars $\eta_i\ (i=1\,,2)$, a singly-charged $SU(2)_L$ singlet scalar
$\chi^+$, and a neutral $SU(2)_L$ singlet scalar $\chi^0$ to the standard model.
Due to two inert $SU(2)_L$  doublets, a single $N_R$ is enough to obtain the observed lepton mixing as well as neutrino masses~\cite{Hehn:2012kz}.
We assume that  the SM-like Higgs $\Phi$ and $\chi^0$ have respectively  vacuum
expectation value (VEV); $v/\sqrt2$ and $v'/\sqrt2$. 
The $Z_2$ symmetry assures the stability of DM ($X_R$) at tree level, but $Z'_2$ is spontaneously broken by $v'$.
As a consequence of the $Z'_2$ breaking, the mixing between $X_R$ and active neutrinos are induced at two loop level.

The relevant renormalizable Lagrangian for Yukawa sector and scalar potential under these assignments
are given by
\begin{eqnarray}
\mathcal{L}_{Y}
&=&
(y_\ell)_a \bar L_{La} \Phi e_{Ra} + (y^i_{\eta})_a \bar L_{La}  N_R \eta^*_i + (y_{\chi})_a  \bar X^c_R e_{Ra} \chi^{+} 
+M_{N}\bar N^c_R N_R 
+M_{X}\bar X^c_R X_R 
+\rm{h.c.} \label{Lag:Yukawa}\\ 
\mathcal{V}
&=& 
 m_1^{2} \Phi^\dagger \Phi + (m_2^{2})_{ij} \eta^\dagger_i \eta_j + m_3^{2} \chi^+ \chi^-  + m_4^{2} |\chi^0|^2  
 + \sqrt2a_{ij} [\chi^0\eta_i^t (i\tau_2) \eta_j \chi^- + {\rm h.c.}] 
 \nn\\
&&
  +\lambda_1 (\Phi^\dagger \Phi)^{2} 
  + (\lambda_2)_{ijk\ell} (\eta^\dagger_i \eta_j)(\eta^\dagger_k \eta_\ell)
  +( \lambda_3)_{ij} (\Phi^\dagger \Phi)(\eta^\dagger_i \eta_j) 
  + (\lambda_4)_{ij} (\Phi^\dagger \eta_i)(\eta^\dagger_j \Phi) 
  \\&&+
  (\lambda_5)_{ij} [(\Phi^\dag\eta_i) (\Phi^\dag\eta_j)+{\rm c.c.}]
+
\lambda_6  (\Phi^\dagger \Phi)(\chi^+ \chi^-)+ (\lambda_7)_{ij}  (\eta^\dagger_i \eta_j) (\chi^+ \chi^-)
+
 \lambda_{8} (\chi^{+} \chi^{-})^2 \nn\\
&&
   +\lambda_9 |\chi^0|^4    +\lambda_{10} (\Phi^\dagger \Phi) |\chi^0|^2
  + (\lambda_{11})_{ij} (\eta^\dagger_i \eta_j) |\chi^0|^2   + \lambda_{12} (\chi^+\chi^-) |\chi^0|^2
 ,\nn
\label{HP}
\end{eqnarray}
where $a=1 \mathchar`-3$, $(i,j)=1\,,2$, and the first term of $\mathcal{L}_{Y}$ can generates the (diagonalized) charged-lepton masses.
Each $M_N$ and $M_X$ includes the mass term after the spontaneous breaking $Z'_2$ symmetry, that is, $y_N v'$ and $y_X v'$.
One of the mass parameters, $a_{ij}v'(\equiv \mu_{ij})$, can be chosen to be real
without any loss of generality by renormalizing the phases to scalar bosons, 
and its index $ij (k \ell)$ runs 1 to 2, where $\mu_{ii}$ vanishes due to the antisymmetric property. Also $\mu_{12}\neq\mu_{21}$ is required, otherwise its term vanishes.
The couplings
$\lambda_1$, $(\lambda_2)_{iiii}\ (i=1\,,2)$, $\lambda_{8}$, and  $\lambda_{9}$ have to be positive to stabilize the Higgs potential.
The scalar fields can be parameterized as 
\begin{align}
&\Phi =\left[
\begin{array}{c}
\phi^+\\
\phi^0
\end{array}\right],\
\eta_i =\left[
\begin{array}{c}
\eta^+_i\\
\eta^0_i
\end{array}\right],    \ (i=1\,,2).
\label{component}
\end{align}
The neutral components of the above fields and the singlet scalar fields can be expressed as
\begin{align}
\phi^0=\frac1{\sqrt2}(v+\varphi),\ 
\eta^0_i=\frac1{\sqrt2}(\eta_{Ri}+i\eta_{Ii}),\ 
\chi^0 = \frac1{\sqrt2}(v'+\rho+i\sigma),
\label{Eq:neutral}
\end{align}
where $\Phi$ is the SM-like Higgs, and $v$ is its VEV, which is related to the Fermi constant $G_F$ by $v^2=1/(\sqrt{2}G_F)\approx(246$ GeV)$^2$.
Inserting the tadpole condition about $\varphi$ and $\rho$,
the resulting mass matrix of the CP-even neutral component, 
 is generally given by $m^{2} (\varphi^2) =   2\lambda_1v^2$, $m^{2} (\rho^2) =   2\lambda_{9}v'^2$, and $m^{2} (\varphi\rho) =   \lambda_{10}vv'$,
 however we assume $\lambda_{10}<<\lambda_{1,9}$ to neglect the mixing.
 Hence $\varphi$ and $\rho$ can be identified as  mass eigenstate.
Due to their mixing terms  $\lambda_{3,4,5,11}$, 
the general expression of the masses for the $\eta_i$ has a complicated form 
so we also assume that these terms are negligible compared to the diagonal terms 
of $(m^2_2)_{11(22)}$ for simplicity. 
As a result, we can identify $\eta^{\pm}_i$, $\eta_{R/Ii}$ ( $i = 1\,,2$) as mass eigenstates hereafter.

\subsection{$S$ and $T$ parameters}
The existence of new scalars ($\eta_i$) implies the additional contribution to 
the $S$ and $T$ parameters~\cite{Barbieri:2006dq, Peskin:1991sw}.
One can evaluate the new contribution as
\bea
{\rm S}_{\mathrm{new}}&=&
\frac{1}{2\pi}\sum_{i=1\,,2}
\int_0^1 dx (1-x)x\ln \left[\frac{x m_{\eta_{Ri}}^2+(1-x)m_{\eta_{Ii}}^2}{m^2_{\eta^+_i}}\right],
\eea
\bea
T_{\mathrm{new}}&=&
\frac{1}{32\pi^2 \alpha_{{\rm em}} v^2} \sum_{i=1\,,2}
\Biggl[ F(m_{\eta^+_i}, m_{\eta_{Ri}})+  F(m_{\eta^+_i}, m_{\eta_{Ii}})- F(m_{\eta_{Ii}}, m_{\eta_{Ri}}) \Biggr],
 \\
F(m_1, m_2)&=&
\frac{m_1^2+m_2^2}{2}-\frac{m_1^2m_2^2}{m_1^2-m_2^2}\ln\left(\frac{m_1^2}{m_2^2}\right).
\eea
where $\alpha_{\mathrm{em}}=1/137$ is the fine structure constant.
The allowed deviations from the SM predictions,  under 
$m_{\phi^0}=126$ GeV, are estimated as \cite{Baak:2012kk}
\be
S_{\mathrm{new}}=0.03\pm 0.10,\quad {\rm T}_{\mathrm{new}}=0.05\pm 0.12...
\label{STallowed}
\ee
Assuming $1\leq m_1/m_2\lesssim3$, we can approximately obtain the following constraint  from the $T_{\mathrm{new}}$,
\begin{equation}
\sum_{i=1\,,2}\left(m_{\eta^+_i}-m_{\eta_{Ri}}\right)
\left(m_{\eta^+_i}-m_{\eta_{Ii}}\right)\lesssim133^2~\mathrm{GeV}.
\label{st-cond}
\end{equation}
 Considering these constraints for $S_{\mathrm{new}}$ and $T_{\mathrm{new}}$, we can obtain $\eta^\pm_i \lesssim$ 200 GeV fixing $\eta_{Ri/Ii}\simeq$ 100 GeV.


\subsection{Neutrino mass matrix}
The neutrino mass matrix can be obtained at one-loop level as follows~\cite{Ma:2006km, Hehn:2012kz}:
\begin{figure}[cbt]
\begin{center}
\includegraphics[scale=0.7]{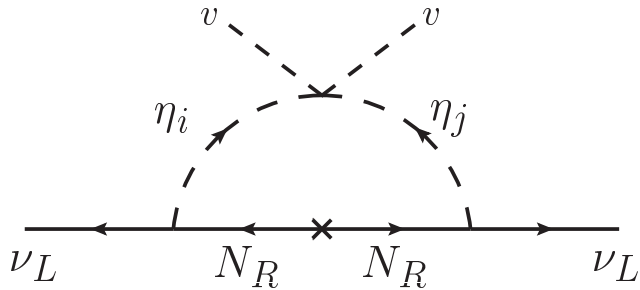}
   \caption{Radiative generation of neutrino masses.  }
   \label{neutrino-diag}
\end{center}
\end{figure}
\bea
({\cal M})_{ab}\approx
-
\frac{(y^i_{\eta})_{a}(y^j_{\eta})_{b}(\lambda_5)_{ij} v^2}{8\pi^2 M_N}
\left[\frac{m^2_{\eta_j}}{m^2_{\eta_i}-m^2_{\eta_j}}\ln\frac{m^2_{\eta_i}}{m^2_{\eta_j}}+ \ln\frac{m^2_{\eta_i}}{M^2_N} \right],
\eea
where we define $m^2_{\eta_{i/j}} \equiv (m^2_{\eta_{Ri/j}}+m^2_{\eta_{Ii/j}})/2, (i,j=1\,,2)$, 
and assume $M_N \gg m^2_{\eta_i},\ m^2_{\eta_j}$ for our convenience to analyze DM. 
In this form, observed neutrino mass differences and their mixings are obtained~\cite{Hehn:2012kz}.
We show a benchmark point to satisfy the data of neutrino masses reported by Planck data~\cite{Ade:2013lta} $\sum m_{\nu}<0.933~\mathrm{eV}$,
as follows:
\begin{align}
&(y_{\eta}) \approx5\times10^{-3},\quad m_{\eta_i}\approx100\ {\rm GeV},\quad m_{\eta_j}\approx110\ {\rm GeV},\quad \lambda_5\approx 10^{-5},
\quad M_N\approx500\ {\rm GeV} .\label{Eq:param_mass}
\end{align}
Then we can obtain the neutrino mass as
\begin{align}
(m_\nu)_{ab} \approx 0.89\ {\rm eV}.
\end{align}
It is worth mentioning that a lepton flavor violating process such as $\mu\to e\gamma$, which provides us the most stringent constraint, should be taken into account.
Applying our benchmark point in Eq.~(\ref{Eq:param_mass}) with $\eta^\pm_i \approx$ 200 GeV (that comes from $S$ and $T$ parameters),  
our branching ratio of $\mu\to e\gamma$ can be estimated as ${\cal O}(10^{-14})$, which is below the experimental bound by MEG $5.7\times10^{-13}$~\cite{meg2}.


\section{Dark Matter}
\label{sec:DM}
We discuss the DM candidate to explain the X-ray line signal reported by~\cite{Bulbul:2014sua,Boyarsky:2014jta}.
On the other hand, our model has two DM candidates; $\eta_{i/j}$ and $X_R$.
To achieve it, $X_R$ has to be dominant number density among these fields.
Here we assume the  mass hierarchy $M_X<m_{\eta_{i/j}}(< N_R)$
\footnote{Since $N_R$ has an accidental symmetry $Z''_2$, it can be also the DM candidate in general.
But we assume $N_R$ can decay enough earlier than the others through the coupling $y_\eta$, that is, $N_R$ is out of DM.}
.
Although a lighter state of $\eta_{i/j}$ would be a DM candidate, 
it cannot be a dominant component of the universe.
This is because the decay channel $2\eta_{i/j}\to 2Z$ opens and its relativistic cross section is enhanced,
If we put its mass to be larger than the mass of neutral gauge boson 
$Z \approx 90$ GeV (but not larger than 500 GeV)~\cite{Hambye:2009pw}.
Notice here that our benchmark point can be always avoidable from the constraint of direct detection search experiments such as LUX~\cite{lux},
since our mass difference between $\eta_{Ri/j}$ and $\eta_{Ii/j}$,
(which is proportional to $\sqrt{\lambda_5} v\approx{\cal O}(1)$ GeV),
is enough large compared to ${\cal O}(100)$ keV that is the lowest bound to escape the constraint of the  inelastic scattering  process via $Z$-boson
for the direct detection search.
Hence we set these masses of  $\eta_{i/j}$ are ${\cal O}(100)$ GeV.
As a result, only $X_R$ can be the DM candidate that is a long-lived particle.

\subsection{Mixing between $X_R$ and $\nu$}
\begin{figure}[cbt]
\begin{center}
\includegraphics[scale=0.7]{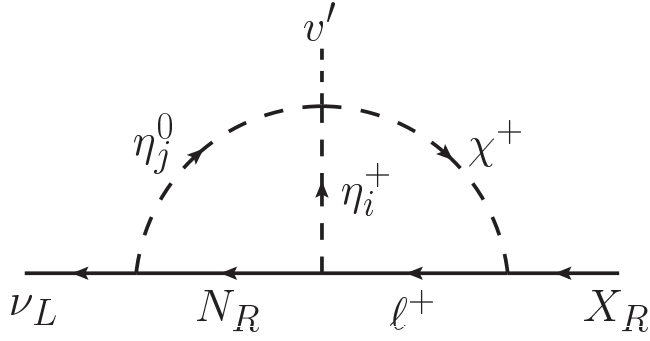}
   \caption{Mixing of $X_R$ with the ordinary neutrinos.}
   \label{neutrino-diag}
\end{center}
\end{figure}

The mixing mass term between $X_R$ and $\nu_a$ is obtained at two-loop level as depicted in Fig.~\ref{neutrino-diag}.
This mass matrix can be given as~\cite{Okada:2014vla} 
\begin{align}
(m_{\nu \mathchar`- X})_{a}
&\approx
\frac{\mu_{ij}}{2}
\left[ \frac{(y_\chi)_{a}(m_\ell)_{a}(y^i_{\eta})^*_{a}
(y^j_{\eta})_{a}}{\sqrt2(4\pi)^4M_N
}
\right]
F\left(\frac{m_{\eta^+_i}^2}{M_N^2}, \frac{m_{\chi^+}^2}{M_N^2},\frac{m_{\eta_j}^2}{M_N^2}\right)+(i\leftrightarrow j),
\label{eq:neutrinomass}
\end{align}
where $m_\ell$ is the SM charged leptons, and the loop function $F$ is computed by
\begin{align}
F\left(X_1,X_2,X_3\right) 
&=
\int_0^1dx\int_0^1dy \int_0^1dz\delta(x+y+z-1)\int_0^1d\alpha\int_0^1d\beta \int_0^1d\gamma\delta(\alpha+\beta+\gamma-1)
\nn\\&\times 
\frac{1}{(z^2-z)(\alpha+\beta X_3 - \gamma \Delta)},
\end{align}
with
\begin{align}
\Delta=\frac{x\frac{(m_{\ell})^2_a}{M_{N}^2}+yX_2+zX_1}{(z^2-z)}.
\end{align}
The mixing between active neutrino and DM is given as
\begin{align}
\theta\equiv \frac{m_{\nu-X}}{M_X} \approx 
5\times10^{-6} \times
\left( \frac{7.1 {\rm keV}}{M_X} \right)
\left( \frac{\mu_{ij}}{100~{\rm GeV}} \right) \left( \frac{y_\chi}{\mathcal{O}(1)} \right) \left( \frac{F}{\mathcal{O}(0.1)} \right)
\label{eq:mixing},
\end{align}
where $M_X$ is the mass of DM and 
$\mu_{ij}$, $y_\chi$, and $F$ are the parameters in our model appeared in Eq. (\ref{eq:neutrinomass}).
The other parameters are already fixed in order to explain the observed neutrino masses 
as in Eq.~(\ref{Eq:param_mass}).
The value $5\times10^{-6}$ is 
an expected mixing angle from the X-ray experiment~\cite{Bulbul:2014sua,Boyarsky:2014jta}.
$F\approx {\cal O}(0.1)$ can be always achieved by fluctuating (almost) the free two mass parameters $m_{\chi^+}$ and $M_N$. 

\subsection{Production of $X_R$}
Here we briefly comment on the production of DM.
We have a scalar field $\chi^0$, which have the Yukawa interaction with $X_R$ in our model.
In this case, the decay of this scalar field into $X_R$ can create the correct abundance of DM (for instance, see Ref.~\cite{Allison:2012qn}).
There might be another possibility.
As we show in the Lagrangian (\ref{Lag:Yukawa}), 
$X_R$ have the Yukawa interaction with a heavy particle $\chi^+$
and $\mathcal{O}(1)$ coupling.
$\chi^+$ is thermalized through the quartic coupling via the SM-like Higgs particle. 
After decoupling $\chi^+$ from the thermal bath, $X_R$ can be created through the decay of $\chi^+$.
Since the amount of the relic abundance of DM depends on 
the decoupling temperature of $\chi^+$, 
that is, the detail magnitude of the scalar coupling constants should be determined.
Such analysis is however beyond the scope at this moment.
Here we would like to note that the DM abundance can be explained 
without conflicting with the results of the current experiments.

From the discussion of the beginning of this section, 
neutral additional particles except $X_R$ decay into the other particles in the early universe.
Therefore, $X_R$ is only the DM candidate in our model.


\section{Conclusions}
We have investigated the radiative neutrino model, in which 
$7.1~{\rm keV}$ X-ray line reported by the analysis 
of the X-ray observation experiments can be successfully realized at the same time.
We have introduced two gauge singlet fermions and three additional inert scalars into the SM
with $Z_2\times Z'_2$ symmetry.
The observed neutrino masses are radiatively generated at the one-loop level, 
and the mixing angle with the ordinary neutrinos is given at the two-loop level.
As a result, the desired mixing $\theta\approx 5\times10^{-6}$ can be naturally explained as well as the neutrino masses.



\end{document}